\documentclass[12pt]{article}
\usepackage{graphicx}
\usepackage{cite}
\usepackage{amsmath,amssymb}
\usepackage[english]{babel}
\usepackage[a4paper, left=3cm, right=2cm, top=2.5cm, bottom=2.5cm]{geometry}

\newcommand{\affiliation}[1]{\begin{center} {\it {#1}} \end{center}}

\newcommand{\Ai}{\mathop{\mathrm{Ai}}}
\newcommand{\Bi}{\mathop{\mathrm{Bi}}}

\begin{document}
\large
\title{Partial electron localization in a finite-size superlattice placed in an electric field}
\author{K.~R.~Vlasov, M.~A.~Pyataev, A.~V.~Shorokhov}

\maketitle

\affiliation{Ogarev Mordovia State University, 430005, Bolshevistskaya, 68, Saransk, Russia}

\begin{abstract}
{Partial electron localization in a finite-size superlattice placed in an electric field is considered. 
The role of electric field in forming of quasilocalized states is investigated. 
A quantitative criterion for the degree of partial localization is suggested
based on analysis of maximal probability density of finding an electron at a given point. 
It is found that with increase in the electric field the degree of localization does not increase 
monotonically. Furthermore, the localization is affected stronger by the amplitude 
of superlattice potential than by the electric field.}
\end{abstract}

\section{Introduction}

The problem of electron localization in a periodic potential under applied 
homogeneous electric field $F$ attracts the scientific interest 
for many years since 1960 when G.H. Wannier suggested the appearance 
of discrete energy spectrum for such a system \cite{Wannier1960}.
Using a basis of Bloch functions and expansion of individual 
terms of the equation in a powers series, 
he concluded that the electron energy spectrum for the system consists of 
a set of equidistant eigenvalues separated by the energy $eFd$, 
where $e$ is the electron charge and $d$ is the period of the potential. 
This phenomenon was called Wannier-Stark localization. 
Later, the arguments of Wannier have been criticized several 
times \cite{Zak1968,Zak1991,Avron1977}. 
In particular, it was proven \cite{Avron1977} 
that energy spectrum of Bloch electron in 
an external electric field is continuous 
for physically meaningful regular potentials. 
Thus, the complete Wannier-Stark localization is impossible. 
However, Wannier-Stark localization can manifest itself as a set of
quasibound states or resonances \cite{Nenciu1991,Gluck2002PRB,Gluck2002}. 
These resonances appear in the energy dependence 
of various physical quantities. They have been observed 
in many experiments and interpreted as evidence 
of Wannier-Stark localization \cite{Mendez1988, Voisin1988, Mendez1993, Tackmann2011}.
Nowadays the concept of Wannier-Stark ladder still attracts much attention because 
it is applied for describing various phenomena in many novel systems 
such as graphene \cite{Ferreira2011, Kelardeh2014},
optical lattices \cite{Beaufils2011, Maury2016}
natural SiC superlattices \cite{Sankin2002,Sankin2011,Sankin2012} and others.

It is obvious that from practical point of view the problem 
of partial electron localization in electric field is more important for the finite 
superlattice because every real structure has finite size.

\section{Hamiltonian and wave function}

\begin{figure}
\begin{center}
\includegraphics[width=0.7\textwidth]{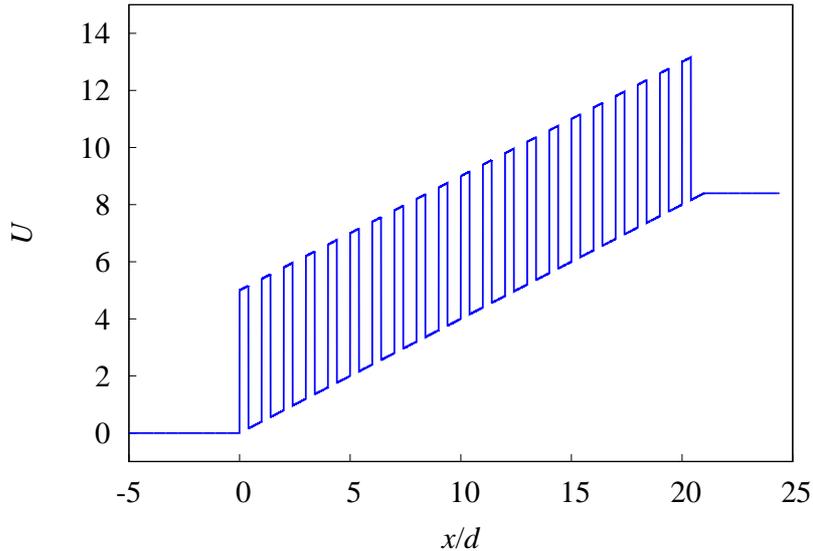}
\caption{\label{VPS17-f1}
The potential energy for the superlattice in the electric field
for $N=20$, $U_0=5 E_d$, $eFd=0.4 E_d$, $a=0.4d$.}
\end{center}
\end{figure}

The aim of the present work is to study the electron localization effect 
in a one-dimensional periodic system containing a few 
dozens of periods, placed in a homogeneous electric field. 
The superlattice in an electric field is described by the Hamiltonian 
\begin{equation}
\hat{H}=\frac{\hat{p}_x^2}{2 m^*}+U(x)-eFx,
\end{equation}
where $m^*$ is the electron effective mass, 
$\hat{p}_x$ is the electron momentum operator
and $U(x)$ is the stepwise superlattice potential given by the following equation:
\begin{equation}
U(x)=\left\{\begin{array}{ll}
U_0, & x-jd<a,\\
0, & x-jd>a.\\
\end{array}\right.
\end{equation}
Here $a$ is the barrier width and $j=[x/d]$ is the integer part of the number $x/d$.
Schematic representation of potential energy for the superlattice in the electric field 
is shown in Fig.~\ref{VPS17-f1}. We assume that electric field vanishes outside 
the superlattice that is why the potential is taken to be constant in regions $x<0$ and $x>Nd$. 

Since we are focused on the role of electric field in formation of quasibound states 
we consider the boundary conditions which do not lead to appearance of 
the discrete energy spectrum in the absence of the electric field.
In contrast to the problem of finding the eigenvalues for the Hamiltonian which takes place 
in case of rigid wall boundary conditions \cite{Ivanov2015}
we consider the scattering problem in which an electron wave of unit amplitude 
goes from the region of zero potential energy ($x\to-\infty $) and then propagates through 
the superlattice or is reflected back (Fig.~\ref{VPS17-f1}). 
It should be noted that propagation is possible only for energies $E>NeFd$
while in the opposite case electrons are completely reflected.

The system can be characterized by three independent parameters with the dimension of energy:
the height $U_0$ of the potential barrier between the layers of superlattice,
the size quantization energy $E_d=\hbar^2/(2 m^* d^2)$, 
and the step of the Wannier-Stark ladder $E_{F}=eFd$.

The electron wave function for $j$-th region can be represented in the form \cite{Liu1986,Brennan1987}
\begin{equation}
\psi_j(x)=\alpha_j \Ai(\xi_j)+\beta_j \Bi(\xi_j),
\end{equation}
where $\Ai(z)$ and $\Bi(z)$ are the Airy functions of the first and second kind respectively, 
$\alpha_j$ and $\beta_j$ are some coefficients. The argument $\xi_j$ of the Airy functions is
given by  
\begin{equation}
\xi_j(x)=\left(\frac{2m^*}{\hbar^2 e^2 F^2}\right)^{1/3}\left(eFx-E+U_j\right),
\end{equation}
where $U_j=0$ for even $j$ and $U_j=U_0$ for odd $j$.

The wave functions at neighbouring regions are related to each other via 
continuity boundary conditions for the wave function and its derivative.
We can get the reciprocal relation from the conditions for the coefficients $\alpha_j$ and $\beta_j$
\begin{equation}
\left(\begin{array}{c}
\alpha_{j+1}\\
\beta_{j+1}
\end{array}\right)
=M_{j+1}^{-1}(x_{j+1})M_{j}(x_{j+1})
\left(\begin{array}{c}
\alpha_{j}\\
\beta_{j}
\end{array}\right),
\end{equation}
where $x_j$ is the point separating $(j-1)$-th and $j$-th regions. 
The matrix $M_j(x)$ has the form 
\begin{equation}
\label{Mj}
M_j(x)
=\left(\begin{array}{cc}
\Ai(\xi_j) & \Bi(\xi_j)\\
\Ai'(\xi_j) & \Bi'(\xi_j)
\end{array}\right).
\end{equation}

The wave function in the region $x<0$ is a superposition of incident and reflected waves 
\begin{equation}
\label{psi0}
\psi_0(x)
=\exp(ik_0 x)+r\exp(-ik_0 x),
\end{equation}
where $r$ is the reflection amplitude and $k_0=(2 m^* E)^{1/2}/\hbar$.
The region $x>Nd$ contains only propagated wave with the amplitude $t$ 
\begin{equation}
\label{psin}
\psi_{n}(x)
=t\exp(ik_n x),
\end{equation}
where $n=2N+1$ and 
\begin{equation}
\label{kn}
k_n=\frac{\sqrt{2 m^* (E-NeFd)}}{\hbar}.
\end{equation}
We note that equation (\ref{psin}) is valid in both cases $E>NeFd$ and $E<NeFd$. 
However, in the second case, the wave vector $k_n$ has only imaginary part  
and $t$ can not be regarded as a transmission amplitude.

From equations (\ref{Mj}), (\ref{psi0}) and (\ref{psin}) 
we can get the following relation between the reflection amplitude 
$r$ and the transmission amplitude $t$:
\begin{equation}
\label{VPS17-eq-tr}
\left(\begin{array}{c}
t\\
0
\end{array}\right)
=S
\left(\begin{array}{c}
1\\
r
\end{array}\right),
\end{equation}
where the scattering matrix $S$ can be represented in the form
\begin{equation}
S=L_{n}^{-1}M_{n-1}(x_{n})M^{-1}_{n-1}(x_{n-1})\ldots M_{1}(x_{2})M_{1}^{-1}(x_{1})L_0.
\end{equation}
Here $L_{0}$ is given by 
\begin{equation}
L_0
=\left(\begin{array}{cc}
1 & 1\\
ik_0 & -i k_0
\end{array}\right)
\end{equation}
and $L_{n}$ has the form
\begin{equation}
L_{n}
=\left(\begin{array}{cc}
\exp(ik_n x_n) & \exp(-ik_n x_n)\\
ik_n \exp(ik_n x_n) & -i k_n \exp(-ik_n x_n)
\end{array}\right),
\end{equation}
where $n=2N+1$ and $k_n$ is given by Eq.~(\ref{kn}).
Calculating matrix $S$ and then solving Eq.~(\ref{VPS17-eq-tr})  
we can obtain transmission and reflection amplitudes 
and then find all the coefficients for the wave function.

\section{Results and discussion}

\begin{figure}
\begin{center}
\includegraphics[width=0.7\textwidth]{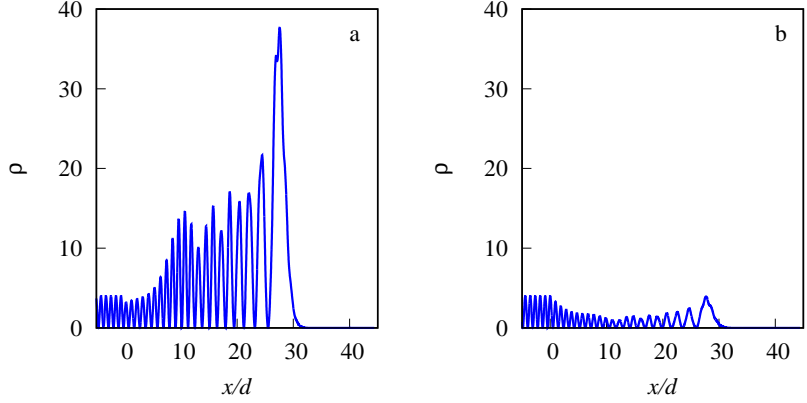}
\caption{\label{VPS17-f2}
Dependence of probability density $\rho$ on electron $x$-coordinate 
for $N=40$, $U_0=2.5E_d$, $eFd=0.4 E_d$
at $E=12.45E_d$ (a) and $E=12.6E_d$ (b).}
\end{center}
\end{figure}

As a criterion of partial electron 
localization we consider the probability density $\rho(x)=|\psi(x)|^2$
of finding an electron at a given point $x$. 
The functions $\rho(x)$ for two different values of energy corresponding 
to total electron reflection are shown in Fig.~\ref{VPS17-f2}. 
The probability density is normalized to unit amplitude of the incident wave.

One can see that depending on energy total electron reflection can be accompanied 
by a significant increase in $\rho(x)$ (Fig.~\ref{VPS17-f2}~a.) 
or can occur without any growth of $\rho(x)$ (Fig.~\ref{VPS17-f2}~b.). 
Relatively small change in energy can lead to a significant change in the  maximal value of $\rho(x)$.  
The large values of $\rho(x)$ can be interpreted 
as long time of electron staying at a given place
and can be considered as an indirect attribute of the partial localization. 

For the further analysis we have investigated the dependence of 
maximal probability density $\rho_{max}$ on the electron energy $E$. 
\begin{figure}
\begin{center}
\includegraphics[width=0.7\textwidth]{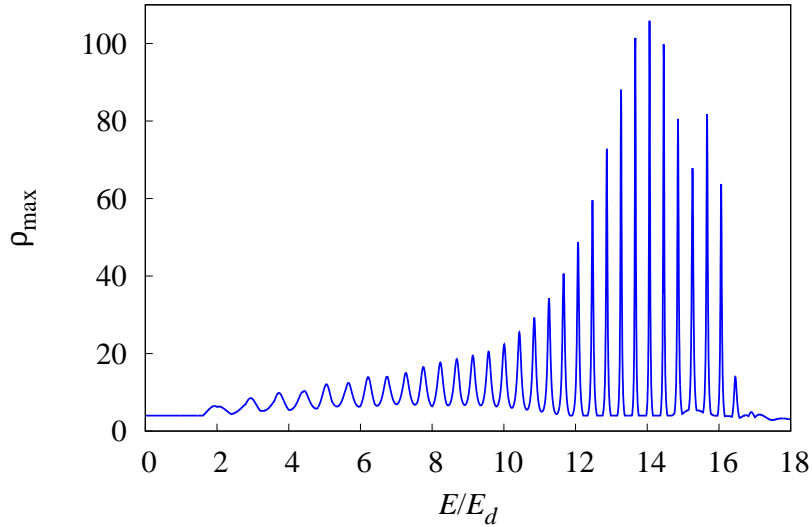}
\caption{\label{VPS17-f3}
Dependence of maximal probability density $\rho_{max}$ on electron energy $E$
for the same parameters as in Fig.~\ref{VPS17-f2}.}
\end{center}
\end{figure}
One can see from Fig.~\ref{VPS17-f3} that 
in  certain intervals of energy $\rho_{max}$ oscillates as a function of $E$. 
The period of the oscillations tends to the distance $eFd$ between the Stark levels
when the energy approaches the value $NeFd$. 
The amplitude of the oscillations can be used as the criterion of partial localization. 
We note that in the case of total reflection the probability density 
always reaches the value of $4$ if the amplitude of incident wave equals unity. 
That is why the amplitude of oscillations 
in Fig.~\ref{VPS17-f3} is determined mostly by the maximal value of peak on the dependence $\rho_{max}(E)$.

\begin{figure}
\begin{center}
\includegraphics[width=0.7\textwidth]{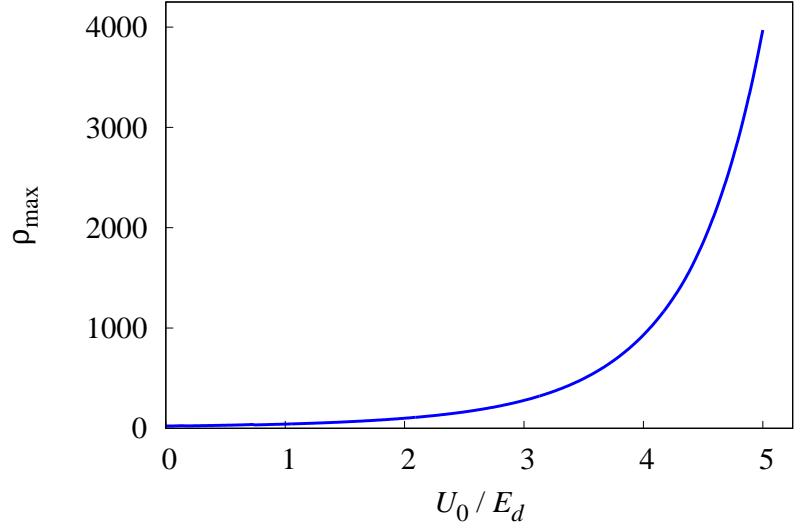}
\caption{\label{VPS17-f4}
Dependence of maximal probability density $\rho_{max}$ on the amplitude $U_0$ 
of the superlattice potential for $N=50$, $eFd=0.3 E_d$ and $a=0.4d$.}
\end{center}
\end{figure}

At the next step we consider the dependence of the maximal oscillation amplitude on 
parameters $F$ and $U_0$. For this reason we found global maximum 
of $\rho$ as a function of $x$ and $E$ for given values of $U_0$ and $F$. 
The dependence of $\rho_{max}$ on the superlattice potential $U_0$ at fixed electric field $F$  
is shown in Fig.~\ref{VPS17-f4}.
The  maximal amplitude of probability density increases very quick with increase in potential barrier $U_0$.
Thus the localization can be clearly seen in the case of high potential barriers (weak coupled superlattices).   

\begin{figure}
\begin{center}
\includegraphics[width=0.7\textwidth]{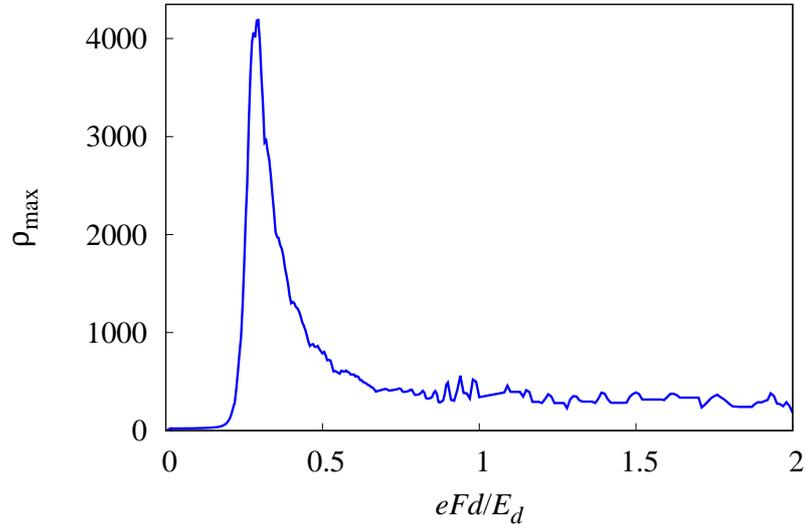}
\caption{\label{VPS17-f5}
Dependence of maximal probability density $\rho$ on the electric field $F$
for $N=50$, $U_0=5E_d$ and  $a=0.4d$.}
\end{center}
\end{figure}

One could expect that the amplitude of oscillations 
should increase monotonically with $F$ 
if the electric field is the main reason of the electron localization. 
However, this is not true. 
The dependence of maximal probability density $\rho_{max}$ on
electric field is shown in Fig.~\ref{VPS17-f5}. 
One can see that the amplitude of probability oscillations increases 
with electric field in low fields but decreases in higher fields. 

\begin{figure}
\begin{center}
\includegraphics[width=0.7\textwidth]{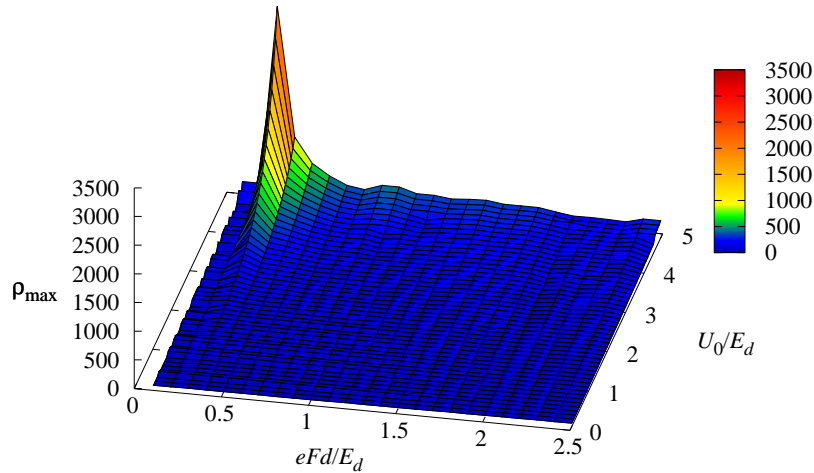}
\caption{\label{VPS17-f6}
Dependence of maximal probability density $\rho_{max}$ on potential barrier height $U_0$ 
and electric field $F$ for $N=50$, $a=0.4d$.}
\end{center}
\end{figure}

Fig.~\ref{VPS17-f6} shows the dependence of $\rho_{max}$ on both parameters $F$ and $U_0$. 
One can see that maximal probability density increases with increase 
in height $U_0$ of the potential barrier. 
At the same time, the dependence of $\rho_{max}$ on $F$ is non-monotonic 
that means that the increase in field does not obligatorily lead to higher degree of localization.

\section{Conclusions}
So we can conclude that in the finite-size superlattice the amplitude $U_0$ of potential difference 
plays more important role in formation of the quasilocalized electron states than the electric field. 
The necessary condition for observation of Wannier-Stark resonances is low coupling between neighbouring 
quantum wells in the superlattice. On the other hand, there is some ``optimal'' electric field for every superlattice 
which makes the resonances most clearly seen. 

\section*{Acknowledgements}
The work has been supported by the RFBR (grant no. 17-02-00969).


\begin{thebibliography}{8}
\itemsep-2pt

\bibitem{Wannier1960}
G.~H.~Wannier, 
{Phys. Rev.} \textbf{117}, 432 (1960).

\bibitem{Zak1968}
J.~Zak, 
{Phys. Rev. Lett.} \textbf{20}, 1477 (1968).

\bibitem{Zak1991}
J.~Zak, 
{Phys. Rev. B} \textbf{43}, 4519 (1991).

\bibitem{Avron1977}
J.~E.~Avron, J.~Zak, A.~Grossmann, and L.~Gunther,
{J. Math. Phys.} \textbf{18}, 918 (1977).

\bibitem{Nenciu1991}
G.~Nenciu,
Rev. Mod. Phys. \textbf{63}, 91 (1991).

\bibitem{Gluck2002PRB}
M.~Gl\"uck, A.~R.~Kolovsky, H.~J.~Korsch, and F.~Zimmer, 
Phys. Rev. B \textbf{65}, 115302 (2002).

\bibitem{Gluck2002}
M.~Gl{\"u}ck, A.~R.~Kolovsky, and H.~J.~Korsch,
Phys. Rep. \textbf{366}, 103 (2002).

\bibitem{Mendez1988}
E.~E.~Mendez, F.~Agullo-Rueda, and J.~M.~Hong, 
Phys. Rev. Lett. \textbf{60}, 2426 (1988).

\bibitem{Voisin1988}
P.~Voisin, J.~Bleuse, C.~Bouche, S.~Gaillard, C.~Alibert, and A.~Regreny, 
Phys. Rev. Lett. \textbf{61}, 1639 (1988). 

\bibitem{Mendez1993}
E.~E.~Mendez and G.~Bastard,
Phys. Today \textbf{46}, 34 (1993).

\bibitem{Tackmann2011}
G.~Tackmann, B.~Pelle, A.~Hilico, Q.~Beaufils, and F.~Pereira dos Santos,
Phys. Rev. A \textbf{84}, 063422 (2011).

\bibitem{Ferreira2011}
G.~J.~Ferreira, M.~N.~Leuenberger, D.~Loss, and J.~C.~Egues,
Phys. Rev. B \textbf{84}, 125453 (2011)

\bibitem{Kelardeh2014}
H.~K.~Kelardeh, V.~Apalkov, and M.~I. Stockman,
Phys. Rev. B \textbf{90}, 085313 (2014).

\bibitem{Beaufils2011}
Q.~Beaufils, G.~Tackmann, X.~Wang, B.~Pelle, S.~Pelisson, P.~Wolf, and F.~Pereira dos Santos,
Phys. Rev. Lett. \textbf{106}, 213002 (2011).

\bibitem{Maury2016}
A.~Maury, M.~Donaire, M.-P.~Gorza, A.~Lambrecht, and R.~Gu{\'e}rout,
Phys. Rev. A \textbf{94}, 053602 (2016).

\bibitem{Sankin2002}
V.~I.~Sankin,
Semiconductors \textbf{36}, 7, 717 (2002).

\bibitem{Sankin2011}
V.~I.~Sankin, A.~V.~Andrianov, A.~O.~Zakhar'in, and A.~G.~Petrov,
JETP Letters \textbf{94}, 362 (2011).

\bibitem{Sankin2012}
V.~I.~Sankin, A.~V.~Andrianov, A.~O.~Zakhar`in, and A.~G.~Petrov, 
{Appl. Phys. Lett.} \textbf{100}, 111109 (2012).

\bibitem{Ivanov2015}
K.~A.~Ivanov, A.~G.~Petrov, M.~A.~Kaliteevski, and A.~J.~Gallant,
{JETP Lett.} \textbf{102}, 796 (2015). 

\bibitem{Liu1986}
W.~W.~Lui and M.~Fukuma,
J. Appl. Phys. \textbf{60}, 1555 (1986).

\bibitem{Brennan1987}
K.~F.~Brennan and C.~J.~Summers,
J. Appl. Phys. \textbf{61}, 614 (1987).


\end{thebibliography}
\end{document}